\PassOptionsToPackage{table}{xcolor}
\ifdefined\pdfobjcompresslevel
\fi
\documentclass[sigconf]{acmart}
\copyrightyear{2026}
\acmYear{2026}
\setcopyright{cc}
\setcctype{by}
\acmConference[RecSys '26]{20th ACM Conference on Recommender Systems}{September 27-October 02, 2026}{Minneapolis, MN, USA}
\acmBooktitle{20th ACM Conference on Recommender Systems (RecSys '26), September 27-October 02, 2026, Minneapolis, MN, USA}
\acmDOI{10.1145/3773078.3831883}
\acmISBN{979-8-4007-2284-4/2026/09}
\usepackage{amsfonts}
\usepackage{enumitem}
\usepackage{multirow}
\usepackage{colortbl}
\definecolor{RowBlue}{HTML}{E6F2FF}

\begin{document}

\title{CAPTS: Channel-Aware, Preference-Aligned Trigger Selection for Multi-Channel Item-to-Item Retrieval}

\author{Xiaoyou Zhou}
\authornote{Equal contribution.}
\orcid{0009-0003-0204-1471}
\affiliation{%
  \institution{Kuaishou Technology}
  \city{Beijing}
  \country{China}}
\email{zhouxiaoyou@kuaishou.com}

\author{Yuqi Liu}
\authornotemark[1]
\orcid{0009-0004-6933-0478}
\affiliation{%
  \institution{Kuaishou Technology}
  \city{Beijing}
  \country{China}}
\email{liuyuqi10@kuaishou.com}

\author{Zhao Liu}
\authornotemark[1]
\orcid{0009-0004-7313-3864}
\affiliation{%
  \institution{Kuaishou Technology}
  \city{Beijing}
  \country{China}}
\email{liuzhao09@kuaishou.com}

\author{Xiao Lv}
\authornote{Corresponding author.}
\orcid{0000-0002-3393-4438}
\affiliation{%
  \institution{Kuaishou Technology}
  \city{Beijing}
  \country{China}}
\email{lvxiao03@kuaishou.com}

\author{Bo Chen}
\orcid{0000-0003-3750-2533}
\affiliation{%
  \institution{Kuaishou Technology}
  \city{Beijing}
  \country{China}}
\email{renze03@kuaishou.com}

\author{Ruiming Tang}
\authornotemark[2]
\orcid{0000-0002-9224-2431}
\affiliation{%
  \institution{Kuaishou Technology}
  \city{Beijing}
  \country{China}}
\email{tangruiming@kuaishou.com}

\author{Guorui Zhou}
\orcid{0009-0002-8550-279X}
\affiliation{%
  \institution{Kuaishou Technology}
  \city{Beijing}
  \country{China}}
\email{zhouguorui@kuaishou.com}

\author{Han Li}
\orcid{0009-0000-9801-9292}
\affiliation{%
  \institution{Kuaishou Technology}
  \city{Beijing}
  \country{China}}
\email{lihan08@kuaishou.com}

\author{Kun Gai}
\orcid{0000-0002-3636-3618}
\affiliation{%
  \institution{Unaffiliated}
  \city{Beijing}
  \country{China}}
\email{gai.kun@qq.com}

\renewcommand{\shortauthors}{Zhou et al.}

\begin{abstract}
Large-scale industrial recommender systems adopt multi-channel retrieval for candidate generation, combining direct user-to-item (U2I) retrieval with two-hop user-to-item-to-item (U2I2I) pipelines. In U2I2I, the system selects a small set of historical interactions as triggers to seed item-to-item (I2I) retrieval across multiple channels. In production, triggers are often selected using rule-based policies or learned scorers and tuned channel by channel. However, these practices face two challenges: biased value attribution, which values triggers by on-trigger feedback rather than downstream retrieval utility, and uncoordinated routing, where channels independently select triggers under a shared quota, increasing cross-channel overlap. To address these challenges, we propose Channel-Aware, Preference-Aligned Trigger Selection (CAPTS), a framework that treats multi-channel trigger selection as a learnable routing problem. CAPTS introduces a Value Attribution Module (VAM) that credits each trigger with subsequent engagement from items retrieved through each I2I channel, and a Channel-Adaptive Trigger Routing (CATR) module that coordinates trigger-to-channel assignment. Offline experiments and large-scale online A/B tests on Kwai, Kuaishou's international short-video platform, show that CAPTS consistently improves multi-channel recall offline and delivers +0.713\% total app time spent and +0.586\% average app time spent per device online.
\end{abstract}



\maketitle


\begin{figure}[t]
  \centering
  \includegraphics[width=\linewidth]{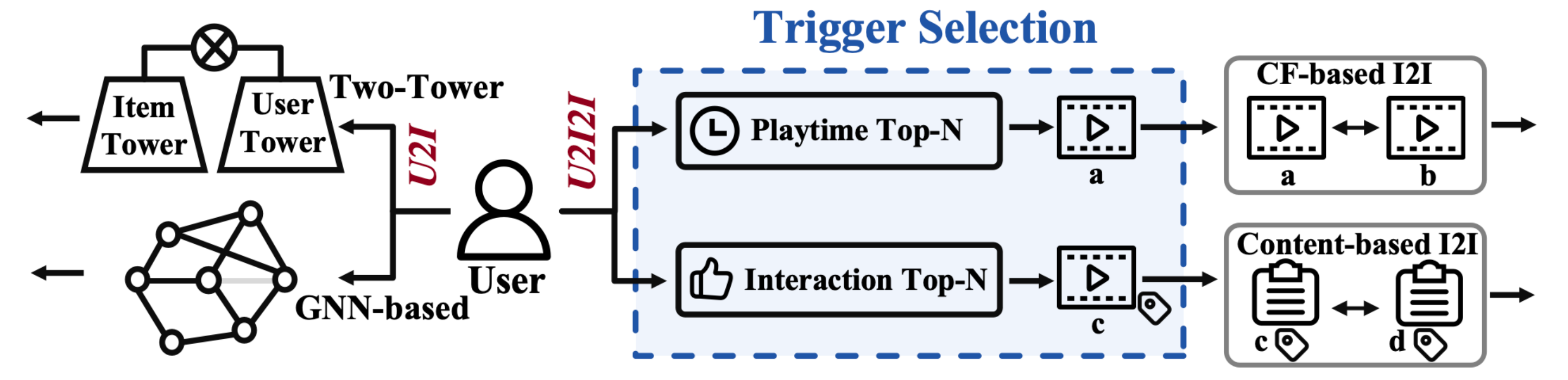}
  \caption{Multi-channel retrieval with two candidate-generation paradigms: direct U2I and two-hop U2I2I. 
We study trigger selection in U2I2I, where a small set of history items is selected to seed downstream I2I retrieval channels. }
  \Description{A user history feeds two parallel candidate-generation paths. The direct U2I path retrieves items from the user representation, while the U2I2I path first selects a small set of historical trigger items and then sends them to multiple item-to-item retrieval channels.}
  \label{fig:u2i2i}
\end{figure}

\section{Introduction}\label{sec:intro}

Industrial recommender systems~\cite{youtube, netflix} typically follow a cascaded pipeline of retrieval, ranking, and re-ranking. Retrieval reduces a corpus of millions to billions of items to a manageable candidate set for downstream scoring, and thus largely determines the ceiling of end-to-end recommendation quality. To broaden interest coverage and improve robustness, production systems commonly adopt multi-channel retrieval, where multiple retrieval channels run in parallel with complementary modeling and retrieval mechanisms~\cite{MIND,ComiRec,KGNN}.

As illustrated in Figure~\ref{fig:u2i2i}, candidate generation in production multi-channel retrieval commonly follows two paradigms: direct user-to-item (U2I)~\cite{U2I_Youtube,U2I_ali} and two-hop user-to-item-to-item (U2I2I)~\cite{PinSage,Pixie,U2I2I_taobao,U2I2I_pinterest}. 
In U2I2I, the system first selects a small set of triggers\footnote{We use the terms interacted item, trigger, and trigger item interchangeably.} from a user’s recent interactions as seeds, and then uses each trigger to retrieve candidates through multiple item-to-item (I2I) channels with different retrieval mechanisms. 
Because triggers serve as the entry points to downstream I2I retrieval, they largely determine the candidate space exposed to these channels, making trigger selection central to multi-channel U2I2I. 
However, much of the existing work~\cite{ULIM,MTMH,Trinity} focuses on the I2I step, leaving trigger selection comparatively underexplored. 
In this work, we focus on trigger selection in multi-channel U2I2I.

In production, triggers are selected from recent interactions~\cite{Pixie,GraphJet,U2I_Youtube} using Watch Time, Liked, or Shared Top-$N$ rules.
Learned trigger scorers (e.g., PDN~\cite{PDN}) are also used, but they are typically developed for a single I2I retrieval channel.
To accommodate heterogeneous I2I channels and practical serving constraints, engineers commonly tune trigger selection in a channel-by-channel manner based on empirical experience.
Online A/B tests refine these choices but leave two challenges.

\begin{figure}[t]
  \centering
  \includegraphics[width=\linewidth]{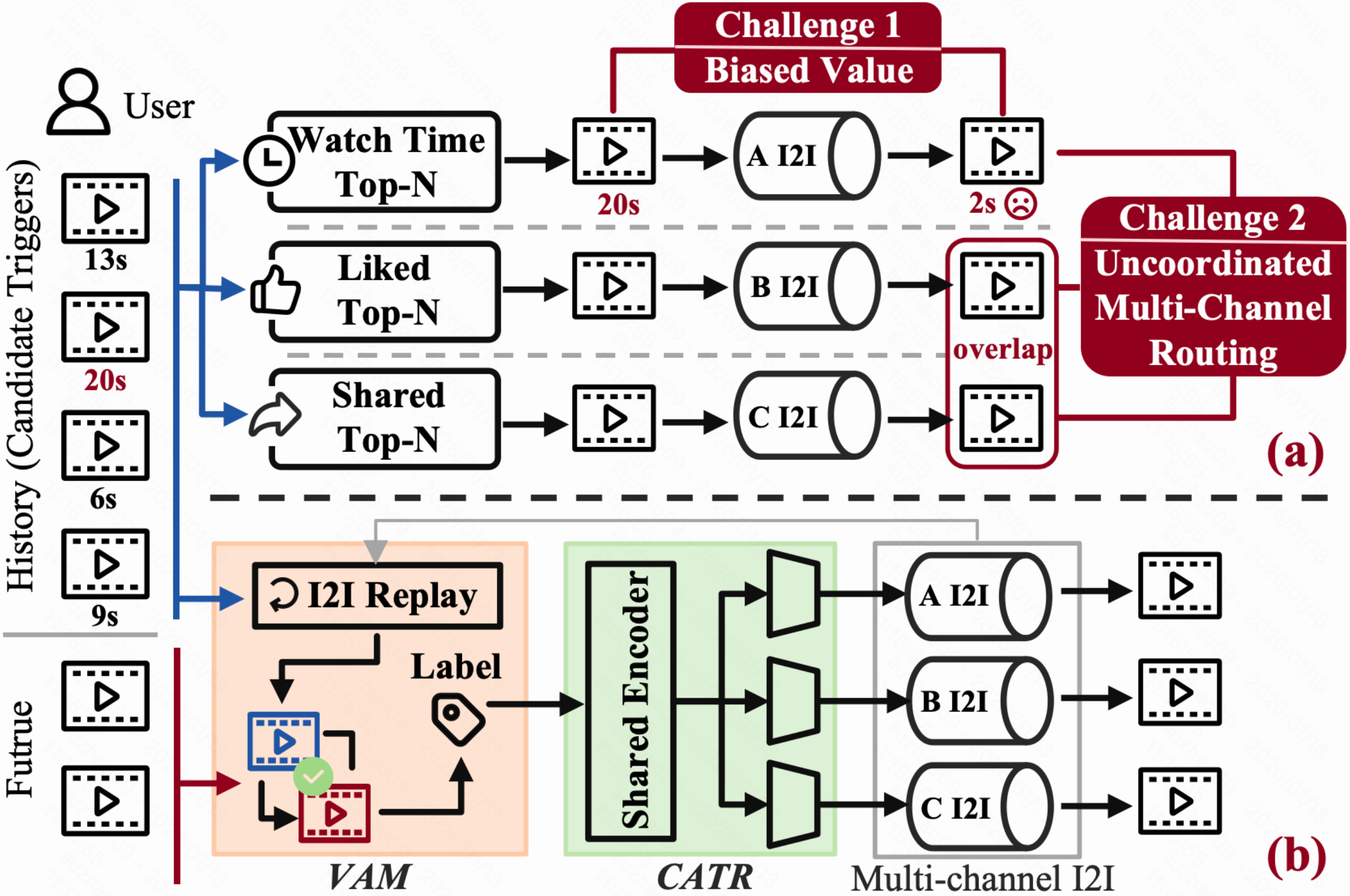}
  \caption{Challenges in multi-channel U2I2I trigger selection and CAPTS.
(a) Conventional trigger selection often relies on direct feedback on the trigger item itself and feeds the selected triggers to multiple downstream I2I channels, which can lead to biased value attribution and uncoordinated multi-channel routing.
(b) CAPTS addresses these issues with two modules: VAM for look-ahead value attribution and CATR for channel-adaptive trigger routing across I2I channels.}
  \Description{Two side-by-side diagrams compare conventional and CAPTS trigger selection. The conventional pipeline scores triggers from direct feedback and routes the same high-scoring items independently to several channels; CAPTS instead attributes downstream value with VAM and coordinates channel-specific routing with CATR.}
  \label{fig:intro}
\end{figure}

As illustrated in Figure~\ref{fig:intro}(a), these practices highlight two persistent challenges in multi-channel trigger selection: biased value attribution and uncoordinated multi-channel routing.
\textbf{First, biased value attribution.} In short-video recommendation, triggers are often selected by simple Top-$N$ rule-based policies based on direct feedback on the trigger item itself, such as watch time and explicit actions (e.g., likes and shares).
However, these signals reflect engagement on the trigger itself and ignore the trigger's downstream utility as a seed.
In U2I2I, a trigger is valuable only insofar as it can bring back useful items via downstream I2I retrieval and drive subsequent engagement on those retrieved items.
When trigger selection is guided solely by feedback on the trigger itself without accounting for this downstream effect, value attribution becomes biased and can yield suboptimal triggers for multi-channel U2I2I.
What we need instead is look-ahead, preference-aligned value attribution that credits a trigger by the subsequent engagement generated by items retrieved from it.
\textbf{Second, uncoordinated multi-channel routing.}
To broaden coverage and improve diversity in candidate generation, modern recommender systems often deploy multiple I2I channels with substantially different retrieval mechanisms, including collaborative filtering, model-based retrieval, and multi-modal content-based retrieval.
Due to this heterogeneity, the same trigger can yield very different downstream utility across channels, making it challenging to decide which triggers are best suited to which channel.
In practice, trigger selection is often optimized on a channel-by-channel basis based on channel-specific metrics and empirical experience, so each channel tends to greedily pick the triggers that look best for itself.
This channel-by-channel optimization makes channels compete for the same limited retrieval quota, which often increases overlap among retrieved candidates and wastes quota on redundant results, thereby reducing the diversity of the joint candidate set.
Trigger selection should therefore be channel-aware and jointly route triggers across channels to maximize the overall downstream utility of multi-channel retrieval.

To address these challenges, we propose \underline{\textbf{C}}hannel-\underline{\textbf{A}}ware,
\underline{\textbf{P}}reference-Aligned \underline{\textbf{T}}rigger \underline{\textbf{S}}election (CAPTS),
a unified and flexible framework for trigger selection in multi-channel U2I2I retrieval.
As illustrated in Figure~\ref{fig:intro}(b), CAPTS comprises two synergistic modules.
\textbf{The Value Attribution Module (VAM)} mitigates biased value attribution by providing look-ahead supervision that credits a trigger according to the subsequent engagement on items retrieved from it on each I2I channel.
\textbf{The Channel-Adaptive Trigger Routing (CATR)} mitigates uncoordinated routing by learning channel-adaptive trigger scores and coordinating trigger-to-channel assignment to maximize the overall value of multi-channel retrieval.
Overall, CAPTS offers a general framework that can jointly optimize trigger selection across heterogeneous I2I channels.
The framework remains lightweight by building channel-wise supervision and a routing scorer around existing I2I retrievers, while leaving retrieval indices and downstream rankers unchanged.

Our main contributions are summarized as follows:
\begin{itemize}[leftmargin=*,nosep]
  \item To the best of our knowledge, we are the first to define the value of a trigger by the subsequent engagement it induces on items retrieved from this trigger in downstream I2I retrieval, and to formulate multi-channel U2I2I trigger selection as a trigger-to-channel routing problem under this downstream-utility objective.
  \item We propose CAPTS, a unified framework for preference-aligned trigger selection in multi-channel U2I2I retrieval, where VAM provides look-ahead, channel-specific supervision from subsequent engagement on retrieved items for value attribution, and CATR learns channel-adaptive value predictions to support channel-aware trigger selection across heterogeneous I2I channels.
  \item We conduct extensive offline experiments and large-scale online A/B tests on Kwai, Kuaishou’s international short-video platform. CAPTS consistently improves multi-channel recall offline and delivers +0.713\% total app time spent and +0.586\% average app time spent per device online, demonstrating its effectiveness in production.
\end{itemize}

\section{Problem Formulation}\label{sec:formulation}

We study trigger selection for multi-channel I2I retrieval.
A request arrives at time $\tau_0$, where the system observes a user $u$,
the user's recent interaction sequence
$\mathcal{H}_u=\big((i_1,\tau_1,d_1,r_1),\ldots,(i_L,\tau_L,d_L,r_L)\big)$,
ordered by timestamp with $\tau_1<\cdots<\tau_L<\tau_0$,
and a set of retrieval channels $\mathcal{C}$ with per-request budgets $B_c$.
Each interaction tuple $(i_k,\tau_k,d_k,r_k)$ records the consumed item $i_k$,
its timestamp $\tau_k$, watch time $d_k$, and an explicit feedback signal $r_k$
such as like, follow, comment, or share.
All items strictly before $\tau_0$ are eligible triggers for this request.

Let $\mathcal{T}_u(\tau_0)=\{\, i_k \mid (i_k,\tau_k,d_k,r_k)\in\mathcal{H}_u,\ \tau_k<\tau_0 \,\}$
denote the set of eligible triggers for this request.
For each channel $c\in\mathcal{C}$, the system selects a trigger subset $S_c \subseteq \mathcal{T}_u(\tau_0)$
with $|S_c|\le B_c$.
The objective is formulated as:
\begin{equation}
\max_{\{S_c\}_{c\in\mathcal{C}}}
\ \sum_{c\in\mathcal{C}} \sum_{t\in S_c} V_c(t)
\quad
\text{s.t. } |S_c|\le B_c \ \ \forall c\in\mathcal{C},
\label{eq:routing}
\end{equation}
where $V_c(t)$ denotes the downstream utility of assigning trigger $t$ to channel $c$.
We instantiate $V_c(t)$ in the Value Attribution Module (Section~\ref{sec:capts_vam}).

\section{The CAPTS Framework}\label{sec:capts}
This section presents CAPTS, our framework for trigger selection in multi-channel I2I retrieval.
We first introduce the Value Attribution Module (VAM) (Section~\ref{sec:capts_vam}), which is built around the downstream utility $V_c(t)$ in Eq.~\eqref{eq:routing} and constructs per-trigger, per-channel supervision signals.
We then describe the Channel-Adaptive Trigger Routing module (CATR) (Section~\ref{sec:capts_catr}), which leverages these signals to learn a channel-adaptive trigger policy for multi-channel trigger selection.
Finally, we discuss the system design for deploying CAPTS efficiently in production (Section~\ref{sec:capts_deploy}).

\begin{figure}[t]
  \centering
  \includegraphics[width=\linewidth]{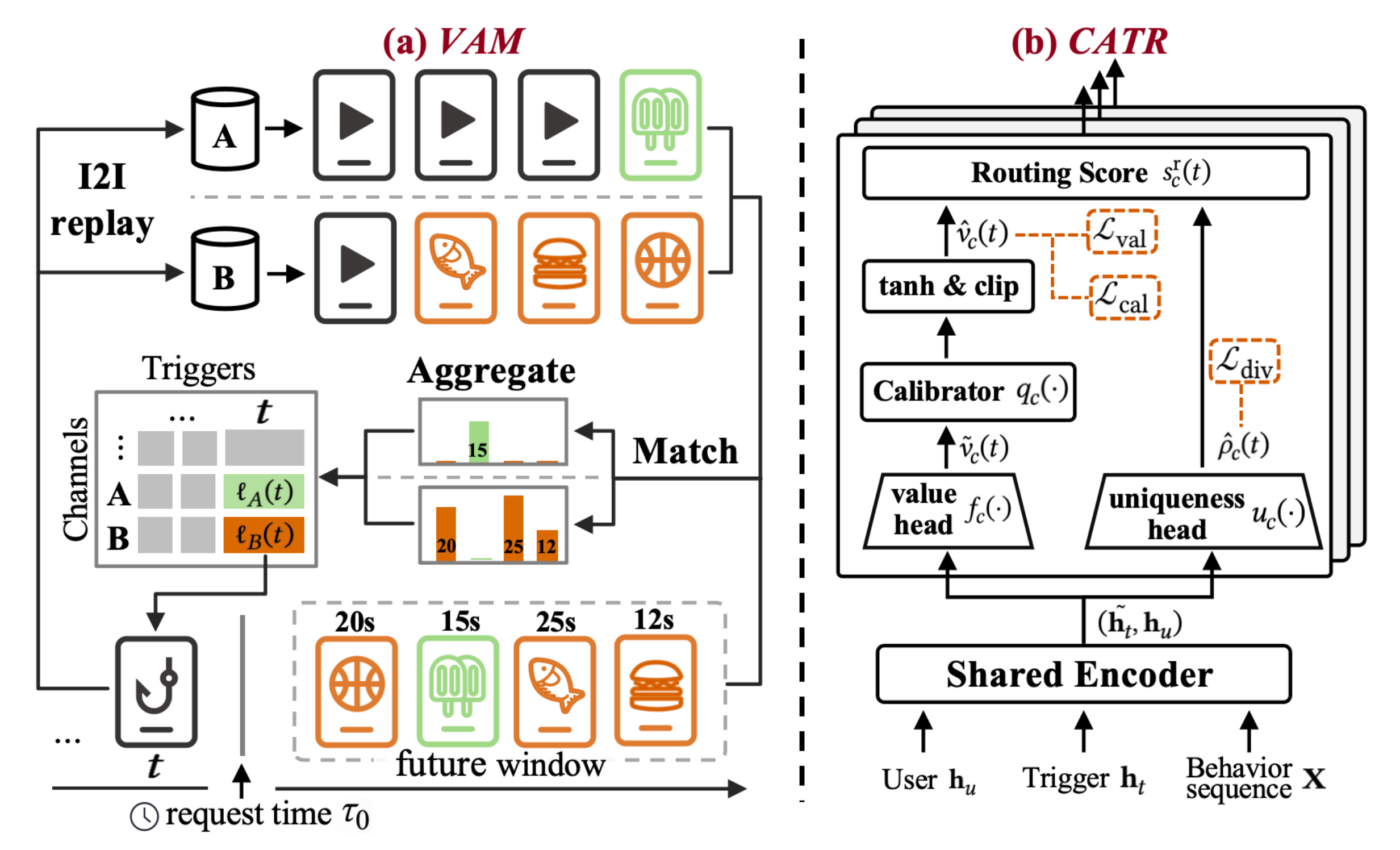}
  \caption{Overview of the CAPTS framework. (a) VAM: at request time $\tau_0$, for each candidate trigger $t$ and retrieval channel $c$, VAM replays I2I retrieval to obtain $\mathcal{R}_c(t,\tau_0)$. It matches retrieved items to a fixed future window $\mathcal{W}_u^{\mathrm{fut}}(\tau_0)$ and aggregates subsequent engagement to form channel-specific supervision for each trigger. (b) CATR: a shared encoder with target attention produces a trigger-aware representation. For each channel, a value predictor with a bounded calibrator estimates the channel-specific trigger value, while a uniqueness head captures cross-channel complementarity. Routing scores combine value and uniqueness for trigger selection.}
  \Description{The VAM panel replays each candidate trigger through every I2I channel, matches retrieved items against a user's future interactions, and aggregates engagement into channel-specific value labels. The CATR panel encodes each trigger, predicts calibrated value and cross-channel uniqueness for every channel, and combines them into routing scores used for channel-specific top-budget selection.}
  \label{fig:vam-catr}
\end{figure}

\subsection{Value Attribution Module (VAM)}\label{sec:capts_vam}

VAM is motivated by a key gap in production trigger selection.
In practice, triggers are often chosen based on direct feedback on the trigger item itself, such as watch time and explicit actions including likes and shares.
Both rule-based policies (e.g., Trinity~\cite{Trinity}) and learned trigger scorers (e.g., PDN~\cite{PDN}) largely follow this paradigm.
However, in U2I2I, a trigger primarily serves as a seed for downstream I2I retrieval, and its usefulness is determined by subsequent engagement on the retrieved items, rather than engagement on the trigger item itself.
Relying only on trigger-level feedback can therefore bias value estimation toward the trigger item and overlook the downstream utility induced by that trigger.
To reduce this bias, we align trigger value with the subsequent engagement on items that are retrieved from the trigger and subsequently consumed, providing preference-aligned supervision for learning trigger value.

To instantiate the downstream utility $V_c(t)$ in Eq.~\eqref{eq:routing}, we define a future window for a logged request instance at time $\tau_0$.
Given a window size $w_s$, let $\mathcal{W}_u^{\mathrm{fut}}(\tau_0)$ denote the set of items consumed in the next $w_s$ effective views after $\tau_0$.
For a candidate trigger $t$ and a retrieval channel $c\in\mathcal{C}$, let $\mathcal{R}_c(t,\tau_0)$ denote the items that channel $c$ would retrieve at time $\tau_0$ when seeded with $t$.
Let $g(j)$ measure the engagement of item $j$, instantiated in this work as watch time on effective views.
We instantiate $V_c(t)$ as the forward value of routing trigger $t$ to channel $c$ as:
\begin{equation}
V_c(t)\triangleq
\mathbb{E}\!\left[
\sum_{j\in \mathcal{R}_c(t,\tau_0)\cap \mathcal{W}_u^{\mathrm{fut}}(\tau_0)} g(j)
\right],
\label{eq:value}
\end{equation}
where the expectation is taken over request instances and the user’s consumption after $\tau_0$.
VAM is designed around this definition to construct per-trigger, per-channel supervision for each request instance, shifting trigger selection from direct feedback on the trigger item to the subsequent engagement induced by its retrieved results.

Figure~\ref{fig:vam-catr}(a) illustrates how VAM constructs per-trigger, per-channel supervision from the user interaction data stream.
For each candidate trigger $t$ and retrieval channel $c$, we replay the timestamp-aligned, channel-specific I2I retrieval seeded by $t$ to obtain the retrieved set $\mathcal{R}_c(t,\tau_0)$.
We instantiate $\mathcal{W}_u^{\mathrm{fut}}(\tau_0)$ from the next $w_s$ effective views after $\tau_0$ and compute engagement only on items that appear in both $\mathcal{R}_c(t,\tau_0)$ and $\mathcal{W}_u^{\mathrm{fut}}(\tau_0)$.
We aggregate engagement over the matched items to define the raw forward reward as:
\begin{equation}
r_c(t,\tau_0) \;=\; \sum_{j\in \mathcal{R}_c(t,\tau_0)\cap \mathcal{W}_u^{\mathrm{fut}}(\tau_0)} g(j),
\label{eq:vam_reward}
\end{equation}
where $r_c(t,\tau_0)$ sums the engagement on items that are both retrieved and consumed within the future window, and $g(j)$ is the watch time on effective views for item $j$.
To mitigate the heavy-tailed scale of aggregated watch time and make supervision stable for learning, we rescale and clip $r_c(t,\tau_0)$ within each channel and obtain a bounded intensity signal as:
\begin{equation}
\ell_c(t,\tau_0) \;=\; \min\!\left\{M_c,\; \max\!\left\{0,\; \frac{r_c(t,\tau_0)}{s_c}\right\}\right\},
\label{eq:vam_label}
\end{equation}
where $s_c>0$ is a channel-specific scale factor and $M_c>0$ caps extreme values. We then construct a binary supervision label for downstream trigger learning as:
\begin{equation}
y_c(t,\tau_0)=\mathbb{I}\{\ell_c(t,\tau_0)\ge \gamma_c\},
\label{eq:vam_binary_label}
\end{equation}
where $\gamma_c\in[0,M_c]$ is a channel-specific threshold defined by the production engagement criterion. VAM outputs the channel-wise supervision vector $\mathbf{y}(t,\tau_0)=\{y_c(t,\tau_0)\}_{c\in\mathcal{C}}$, and uses $\ell_c(t,\tau_0)$ as an intensity-aware signal for sample reweighting and calibration in downstream training.

\subsection{Channel-Adaptive Trigger Routing (CATR)}\label{sec:capts_catr}

Modern candidate generation in production recommender systems commonly deploys multiple heterogeneous I2I retrieval channels in parallel to broaden interest coverage and improve robustness. Multi-channel retrieval increases coverage and diversity, yet it also makes trigger selection harder because the same trigger can yield very different downstream utility across channels. In practice, triggers are often optimized in a channel-by-channel manner based on channel-specific metrics and empirical experience, so each channel greedily selects the triggers that look best for itself without coordination at routing time. Because channels share a limited retrieval quota, such uncoordinated routing tends to concentrate multiple channels on similar triggers, increasing overlap among retrieved candidates and wasting quota on redundant results, which reduces the diversity and overall utility of the joint candidate set. CATR addresses this issue by learning channel-adaptive trigger routing scores and coordinating trigger-to-channel assignment with an explicit complementarity objective, so as to reduce cross-channel redundancy and maximize the overall downstream utility of multi-channel retrieval.

As illustrated in Figure~\ref{fig:vam-catr}(b), CATR uses a shared encoder for user and request context together with lightweight channel-specific heads. For each training instance, the encoder produces a user representation $\mathbf{h}_u$, a trigger representation $\mathbf{h}_t$, and a behavior sequence representation $\mathbf{X}$ from recent interactions. A target-attention module uses $\mathbf{h}_t$ to attend over $\mathbf{X}$ and outputs a trigger-aware summary $\tilde{\mathbf{h}}_t$. For each channel $c$, a channel-specific value head $f_c$ maps $(\tilde{\mathbf{h}}_t,\mathbf{h}_u)$ to a base value prediction as:
\begin{equation}
\tilde{v}_c(t)=\sigma\left(f_c(\tilde{\mathbf{h}}_t,\mathbf{h}_u)\right),
\label{eq:catr_base_value}
\end{equation}
where $c\in\mathcal{C}$ indexes retrieval channels, $f_c(\cdot)$ is the channel-specific value head, and $\sigma(\cdot)$ is the sigmoid activation. Training data are constructed by offline log replay anchored at a historical request time $\tau_0$. For each candidate trigger $t$ and channel $c$, VAM aligns the replayed channel recall with the subsequent consumption window and produces a binary supervision label $y_c(t,\tau_0)\in\{0,1\}$, and it retains the clipped engagement intensity $\ell_c(t,\tau_0)$ as an auxiliary signal for sample reweighting and calibration.

Directly injecting continuous watch-time regression into the main training objective can lead to unstable optimization due to gradient interference across objectives and the strong nonlinear relationship between watch time and user interactions. To incorporate fine-grained intensity signals while keeping probabilistic learning stable, CATR attaches a channel calibrator that performs a bounded correction in the value space as:
\begin{equation}
\hat{v}_c(t)=\mathrm{clip}\!\left(\tilde{v}_c(t)
+\beta\,\tanh\!\big(q_c([\tilde{\mathbf{h}}_t,\mathbf{h}_u,\tilde{v}_c(t)])\big)\right),
\label{eq:catr_calibrator}
\end{equation}
where $q_c(\cdot)$ is the channel-$c$ calibrator network, $\beta$ controls the correction magnitude, and $\mathrm{clip}(\cdot)$ clips the output to $[0,1]$. We fit calibrated values with a weighted binary cross-entropy loss as:
\begin{equation}
\mathcal{L}_{\mathrm{val}}
=-\sum_{c\in\mathcal{C}}\sum_t w_{c,t}\Big[
y_c\log\hat{v}_c+(1-y_c)\log(1-\hat{v}_c)
\Big],
\label{eq:catr_value_loss}
\end{equation}
where $y_c$ and $\hat{v}_c$ abbreviate $y_c(t,\tau_0)$ and $\hat{v}_c(t)$, respectively, and $w_{c,t}=1+\ell_c(t,\tau_0)$ reweights samples by the clipped engagement intensity. To keep calibration stable, we add an intensity-aware calibration loss as:
\begin{equation}
\mathcal{L}_{\mathrm{cal}}
=\sum_{c\in\mathcal{C}}\sum_{t} w^{\mathrm{cal}}_{c,t}\Big\|\hat{v}_c(t)-\frac{\ell_c(t,\tau_0)}{M_c}\Big\|_2^2,
\label{eq:catr_cal_loss}
\end{equation}
where $\ell_c(t,\tau_0)/M_c\in[0,1]$ is the normalized intensity target from VAM, $w^{\mathrm{cal}}_{c,t}=1+\sigma\left(\ell_c(t,\tau_0)\right)$ upweights high-intensity samples, $\sigma(\cdot)$ is the sigmoid function, and $M_c>0$ is the channel-specific clipping cap in Eq.~\eqref{eq:vam_label}.

To reduce cross-channel redundancy and encourage complementarity, CATR augments the value head with a channel-wise uniqueness head. Following VAM, $\mathcal{R}_c(t,\tau_0)$ denotes the retrieved set returned by channel $c$ when seeded by trigger $t$ at time $\tau_0$. The channel-unique subset is defined as:
\begin{equation}
\mathcal{U}_c(t,\tau_0)=\Big\{\,j\in\mathcal{R}_c(t,\tau_0)\;:\; j\notin\mathcal{R}_{c'}(t,\tau_0)\ \text{for all }c'\neq c\,\Big\},
\label{eq:catr_unique_set}
\end{equation}
where $j$ indexes retrieved items. The uniqueness ratio is:
\begin{equation}
\rho_c(t,\tau_0)=\frac{\big|\mathcal{U}_c(t,\tau_0)\big|}{\big|\mathcal{R}_c(t,\tau_0)\big|+\varepsilon},
\label{eq:catr_unique_ratio}
\end{equation}
where $\varepsilon>0$ is a smoothing constant.
We threshold $\rho_c(t,\tau_0)$ to form a binary uniqueness supervision
$y^{\mathrm{u}}_c(t,\tau_0)=\mathbf{1}\{\rho_c(t,\tau_0)>\theta\}$, where $\theta$ is the uniqueness threshold.
A channel-specific uniqueness head predicts the uniqueness ratio as:
\begin{equation}
\hat{\rho}_c(t)=\sigma\left(u_c(\tilde{\mathbf{h}}_t,\mathbf{h}_u)\right),
\label{eq:catr_uniq_head_base}
\end{equation}
where $u_c(\cdot)$ is the channel-specific uniqueness head and $\sigma(\cdot)$ is the sigmoid function.
We train the uniqueness head with the diversity loss as:
\begin{equation}
\mathcal{L}_{\mathrm{div}}
=-\sum_{c\in\mathcal{C}}\sum_t w_{c,t}\Big[
y^{\mathrm{u}}_c\log\hat{\rho}_c
+(1-y^{\mathrm{u}}_c)\log(1-\hat{\rho}_c)
\Big],
\label{eq:catr_div_loss}
\end{equation}
where $y^{\mathrm{u}}_c$ and $\hat{\rho}_c$ abbreviate $y^{\mathrm{u}}_c(t,\tau_0)$ and $\hat{\rho}_c(t)$, respectively; $y^{\mathrm{u}}_c\in\{0,1\}$ is derived from $\rho_c(t,\tau_0)$,
and $w_{c,t}=1+\ell_c(t,\tau_0)$ shares the same intensity-based reweighting as in Eq.~\eqref{eq:catr_value_loss}.
The overall objective is:
\begin{equation}
\mathcal{L}=\mathcal{L}_{\mathrm{val}}+\lambda\,\mathcal{L}_{\mathrm{cal}}+\mu\,\mathcal{L}_{\mathrm{div}},
\label{eq:catr_total_loss}
\end{equation}
where $\lambda$ and $\mu$ control the strengths of calibration and diversity terms.
At inference, CATR forms a diversity-aware routing score as:
\begin{equation}
s^{\mathrm{r}}_c(t)=\hat{v}_c(t)+\eta\,\hat{\rho}_c(t),
\label{eq:catr_route_score}
\end{equation}
where $\eta\ge 0$ trades off value and diversity, and each channel selects its Top-$B_c$ triggers by $s^{\mathrm{r}}_c(t)$ for routing.

\subsection{Production Deployment}\label{sec:capts_deploy}

\begin{figure}[t]
  \centering
  \includegraphics[width=\linewidth]{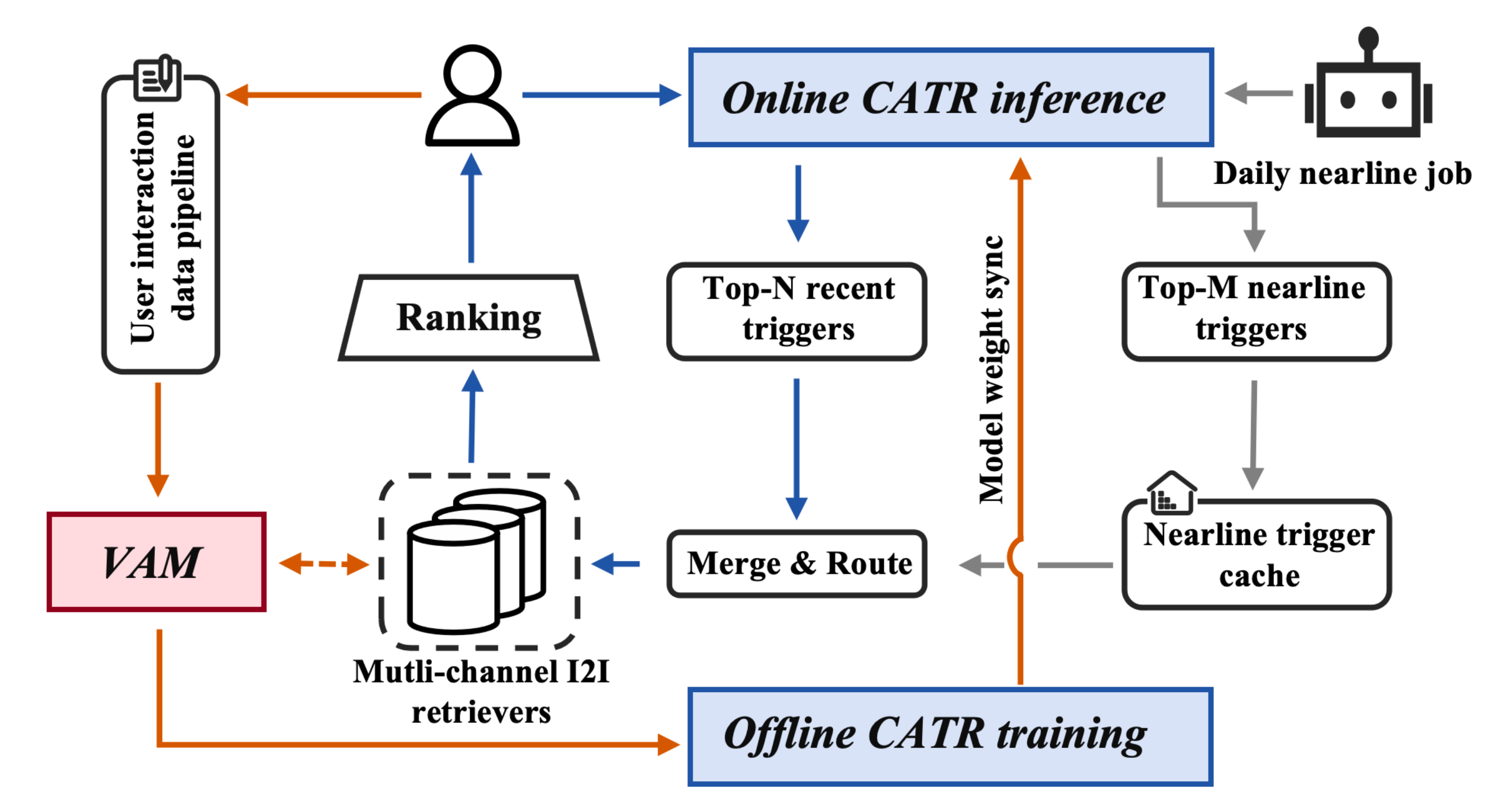}
  \caption{Production deployment of {CAPTS}.
    Orange lines denote offline value attribution and {CATR} training, with periodic model-weight synchronization to online serving.
    Blue lines denote per-request online trigger scoring and routing for multi-channel {I2I} retrieval.
    Grey lines denote daily nearline jobs that refresh the nearline trigger cache for long-term triggers.}
    \Description{A production architecture separates three paths. The offline path builds VAM labels, trains CATR, and periodically synchronizes model weights; the online path scores recent triggers and routes them to multiple I2I channels per request; the nearline path refreshes a cache of long-term triggers that is merged into online routing.}
    \label{fig:deployment}
\end{figure}

In production, trigger scoring must meet strict latency and compute budgets even when the eligible user history is large. To balance responsiveness and interest coverage, CAPTS combines online scoring of recent interactions, typically on the order of $10^2$ to $10^3$ candidates, with a lightweight nearline cache over longer histories, typically on the order of $10^3$ to $10^4$ candidates. The online path focuses on short-term interests available at request time, while the nearline path periodically refreshes high-value long-term triggers during off-peak hours. As shown in Figure~\ref{fig:deployment}, user logs are fed to VAM to construct channel-wise supervision, which is joined with request features to train CATR offline; trained snapshots are then synchronized to online serving. At request time, the service scores recent triggers, merges them with cached long-term triggers, and uses the same CATR routing scores to assign triggers to each I2I channel under per-channel budgets before downstream ranking. This keeps CAPTS localized to trigger selection: online serving adds only lightweight scoring, routing, and cache lookup, while existing I2I retrievers and rankers remain intact.

\section{Experiments}
\label{sec:experiments}

We empirically evaluate CAPTS through the following research questions:

\noindent\textbf{RQ1:} In offline evaluation, does CAPTS outperform strong industrial trigger selection baselines in overall multi-channel recall?\\
\textbf{RQ2:} How much does VAM improve per-channel recall, and how sensitive is it to the future window size?\\
\textbf{RQ3:} Can CATR increase cross-channel diversity and complementarity without degrading overall recall?\\
\textbf{RQ4:} Do the offline gains of CAPTS translate into lifts on key business metrics in large-scale online A/B tests?

\subsection{Offline Experiments (RQ1)}

\subsubsection{Dataset}

\begin{table}[t]
\centering
\caption{Statistics of the industrial short-video dataset.}
\label{tab:dataset}
\renewcommand{\arraystretch}{1.05}
\begin{tabular}{l|l}
\toprule
Metric & Kwai-Industrial \\
\midrule
\#Users                 & 1{,}071{,}280       \\
\#Items                 & 27{,}361{,}781      \\
\#Interactions          & 1{,}230{,}150{,}824 \\
\#Effective views       & 723{,}971{,}024     \\
Avg.\ interactions/user & 1{,}148.3           \\
Avg.\ effective views/user  & 675.8               \\
\bottomrule
\end{tabular}
\end{table}

Offline experiments are conducted on production logs from Kwai, a large-scale international short-video platform.  
We retain users with rich effective-view histories from a recent short-video production snapshot, focusing on cases where trigger selection is needed rather than short histories.
For these users, we collect their historical interactions before the snapshot and construct chronological user histories.  
We use a chronological split: earlier requests and their future-window engagement train VAM, whereas the last 100 effective views are held out exclusively for Recall@K evaluation.
Dataset statistics are summarized in Table~\ref{tab:dataset}.

\subsubsection{Baselines and Metrics}

We compare CAPTS with several industrial trigger selection strategies, including: (i) rule-based trigger selection methods, and (ii) model-based trigger selection methods.
\begin{itemize}[leftmargin=*]
    \item \textbf{TagTop}: A hierarchical popularity baseline that first selects the most exposed tags in the user history and then chooses long-view popular videos within each tag with author de-duplication to keep basic diversity.
    \item \textbf{LTV}: A long-term value strategy that prefers videos whose past exposures lead to sustained follow-up interactions from the same creator or topic, targeting user stickiness rather than short-term clicks.
    \item \textbf{NIC} (New Interest Capture): An in-house strategy for emerging interests that analyzes tag-level statistics, explicitly filters out stable and decaying interests, and uses only videos associated with recently rising tags as triggers.
    \item \textbf{Recent}: A recency-based baseline that selects the most recent effective views as triggers to capture users' short-term interests.
    \item \textbf{PDN}~\cite{PDN}: A path-based deep network that models user $\rightarrow$ trigger $\rightarrow$ target trajectories and jointly learns I2I similarity and trigger importance; we use it only for trigger selection with our fixed production I2I channels.
    \item \textbf{LIC}~\cite{LIC}: A long-term interest clock method for fine-grained time perception in streaming recommendation that derives a time-aware user embedding from long-term behaviors around the current time and scores candidate triggers by matching them with this embedding.
\end{itemize}

To mimic real multi-channel I2I deployment, we replay three production I2I retrieval channels in offline evaluation that cover the major retrieval mechanisms used in our online system: Swing~\cite{U2I2I_taobao} I2I as a collaborative-filtering channel, Marm~\cite{Marm} I2I as a model-based retrieval channel, and MMU~\cite{MMU} I2I as a multi-modal content-based retrieval channel.
All methods use the same retrieval configuration on each channel and differ only in trigger selection.
We use Recall@K as the main metric. 
For multi-channel evaluation, $K$ is the per-channel cutoff: each channel $c$ returns its top-$K$ retrieved candidates $\mathcal{R}_c^K$, and we compute Recall@K on the de-duplicated union $\mathcal{R}_{\mathrm{union}}^K=\bigcup_{c\in\mathcal{C}}\mathcal{R}_c^K$. Due to cross-channel overlap, $|\mathcal{R}_{\mathrm{union}}^K|$ can be smaller than $K|\mathcal{C}|$. 
Recall@K is defined as $\left|\mathcal{W}_u^{\mathrm{fut}}(\tau_0)\cap \mathcal{R}_{\mathrm{union}}^K\right| \big/ \left|\mathcal{W}_u^{\mathrm{fut}}(\tau_0)\right|$, i.e., the fraction of unique videos consumed in the future window that appear in $\mathcal{R}_{\mathrm{union}}^K$.

\subsubsection{Implementation Details}

Unless otherwise stated, we use a future window of size $w_s=100$ effective views. For CATR (Section~\ref{sec:capts_catr}), we set $\beta=0.1$, $\theta=0.8$, and $\lambda=\mu=0.1$.
We empirically set $\eta=0.4$ based on sensitivity analysis, as it yields the best R@2000 and stable performance at smaller retrieval cutoffs.

\subsubsection{Overall Performance}

\begin{table}[t]
\centering
\caption{
Overall multi-channel retrieval performance in offline evaluation. R@K denotes Recall@K with $K \in \{100, 500, 1000, 2000\}$, evaluated on the overall results from three I2I retrieval channels. Underlined numbers indicate the best non-CAPTS baseline at each $K$, including ties. Improv.\ denotes the relative gain of CAPTS over this best baseline, and all gains are statistically significant under paired $t$-tests over requests with $p < 0.05$.
}
\label{tab:overall-offline}
\renewcommand{\arraystretch}{1.05}
\begin{tabular}{l|cccc}
\toprule
Method & R@100 & R@500 & R@1000 & R@2000 \\
\midrule
TagTop & 0.0163 & 0.0659 & 0.1114 & 0.1539 \\
LTV    & 0.0138 & 0.0628 & 0.1117 & 0.1596 \\
NIC    & \underline{0.0204} & 0.0808 & 0.1297 & 0.1719 \\
Recent & \underline{0.0204} & \underline{0.0817} & \underline{0.1335} & \underline{0.1802} \\
\cmidrule(lr){1-5}
PDN    & 0.0102 & 0.0491 & 0.0887 & 0.1252 \\
LIC    & 0.0202 & 0.0793 & 0.1289 & 0.1738 \\
\specialrule{0.08em}{0.35em}{0.35em}
\textbf{CAPTS}   & \textbf{0.0255} & \textbf{0.0963} & \textbf{0.1519} & \textbf{0.1994} \\
\rowcolor{RowBlue}\textbf{Improv.} & +25.0\% & +17.9\% & +13.8\% & +10.7\% \\
\bottomrule
\end{tabular}
\end{table}

Table~\ref{tab:overall-offline} reports the overall offline retrieval performance evaluated on the de-duplicated union of candidates returned by three I2I retrievers: Swing, Marm, and MMU. For each retriever, we take its Top-$K$ retrieved candidates and compute Recall@K on the union after de-duplication. CAPTS achieves the best Recall@K for all $K \in \{100, 500, 1000, 2000\}$ and yields statistically significant improvements over Recent. Recent is our strongest production baseline, reflecting that timeliness is critical for effective candidate generation in short-video retrieval. At $K=2000$, Recent reaches 0.1802, while CAPTS reaches 0.1994, corresponding to a 10.7\% relative improvement. At $K=100$, $500$, and $1000$, CAPTS brings relative gains of 25.0\%, 17.9\%, and 13.8\%, respectively. Overall, these consistent gains align with the design of CAPTS, which aligns trigger scoring with downstream consumption gains and coordinates triggers across channels to reduce redundancy, thereby improving the coverage of the merged retrieval results.

\subsection{VAM Effectiveness (RQ2)}

\subsubsection{Channel-wise Gains of VAM}

\begin{table}[t]
\centering
\caption{
Per-channel Recall@K under isolated channel evaluation on Swing, Marm, and MMU I2I.
Underlined numbers indicate the best non-CAPTS baseline at each $K$.
Improv.\ denotes the relative gain of CAPTS over this best baseline.
}
\label{tab:vam-channel}
\setlength{\tabcolsep}{3.5pt}
\renewcommand{\arraystretch}{1.03}
\begin{tabular}{l|lcccc}
\toprule
Channel & Method & R@100 & R@500 & R@1000 & R@2000 \\
\midrule
\multirow{8}{*}{Swing I2I}
  & TagTop   & 0.0092 & 0.0387 & 0.0687 & 0.1153 \\
  & LTV      & 0.0073 & 0.0352 & 0.0666 & 0.1197 \\
  & NIC      & \underline{0.0119} & 0.0503 & 0.0858 & 0.1306 \\
  & Recent   & \underline{0.0119} & \underline{0.0508} & \underline{0.0880} & \underline{0.1411} \\
  & PDN      & 0.0052 & 0.0263 & 0.0510 & 0.0915 \\
  & LIC      & 0.0118 & 0.0491 & 0.0837 & 0.1338 \\
  \cmidrule(lr){2-6}
  & \textbf{CAPTS}   & \textbf{0.0152} & \textbf{0.0627} & \textbf{0.1068} & \textbf{0.1610} \\
  & \textbf{Improv.} & +27.73\% & +23.43\% & +21.36\% & +14.10\% \\
\midrule
\multirow{8}{*}{Marm I2I}
  & TagTop   & 0.0122 & 0.0493 & 0.0819 & 0.0883 \\
  & LTV      & 0.0106 & 0.0492 & 0.0869 & 0.0936 \\
  & NIC      & 0.0145 & 0.0578 & 0.0891 & 0.0951 \\
  & Recent   & \underline{0.0146} & \underline{0.0596} & \underline{0.0958} & \underline{0.1012} \\
  & PDN      & 0.0083 & 0.0404 & 0.0708 & 0.0751 \\
  & LIC      & 0.0144 & 0.0577 & 0.0935 & 0.0996 \\
  \cmidrule(lr){2-6}
  & \textbf{CAPTS}   & \textbf{0.0163} & \textbf{0.0663} & \textbf{0.1016} & \textbf{0.1040} \\
  & \textbf{Improv.} & +11.64\% & +11.24\% & +6.05\% & +2.77\% \\
\midrule
\multirow{8}{*}{MMU I2I}
  & TagTop   & 0.0009 & 0.0038 & 0.0065 & 0.0108 \\
  & LTV      & 0.0007 & 0.0036 & 0.0070 & 0.0132 \\
  & NIC      & \underline{0.0016} & 0.0057 & 0.0097 & 0.0145 \\
  & Recent   & \underline{0.0016} & \underline{0.0059} & \underline{0.0101} & \underline{0.0169} \\
  & PDN      & 0.0006 & 0.0032 & 0.0062 & 0.0115 \\
  & LIC      & \underline{0.0016} & 0.0057 & 0.0097 & 0.0160 \\
  \cmidrule(lr){2-6}
  & \textbf{CAPTS}   & \textbf{0.0041} & \textbf{0.0138} & \textbf{0.0214} & \textbf{0.0311} \\
  & \textbf{Improv.} & +156.25\% & +133.90\% & +111.88\% & +84.02\% \\
\bottomrule
\end{tabular}
\end{table}

To isolate the impact of VAM across retrieval channels, we conduct single-channel evaluations on Swing, Marm, and MMU I2I. For each channel, we compute per-channel Recall@K using only the candidates retrieved by that channel, while keeping the underlying I2I retriever and retrieval configuration fixed, so that different methods vary only in trigger selection. As shown in Table~\ref{tab:vam-channel}, CAPTS yields consistent improvements across all channels and all $K$ values. At R@2000, CAPTS improves over Recent by 14.10\% on Swing, 2.77\% on Marm, and 84.02\% on MMU. These gains are consistent with the design of VAM, which attributes trigger value to subsequent engagement on items retrieved from the trigger within a future window, thereby mitigating biased value attribution from trigger-side feedback.

\subsubsection{Sensitivity to Future Window}

\begin{figure}[t]
\centering
\includegraphics[width=\linewidth]{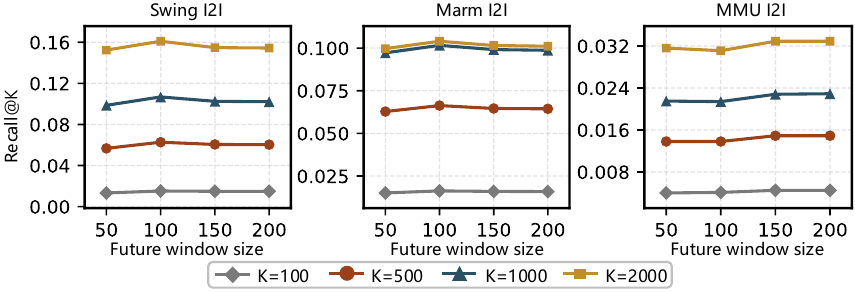}
\caption{Sensitivity of Recall@K to the future window size $w_s$ on Swing, Marm, and MMU I2I. We vary $w_s \in \{50,100,150,200\}$ in the value aggregation model while keeping other training settings fixed, and plot Recall@K for $K \in \{100,500,1000,2000\}$.}
\Description{Three groups of line charts show Recall at 100, 500, 1000, and 2000 as the future-window size increases from 50 to 200 for Swing, Marm, and MMU. Swing and Marm peak or stabilize near a window of 100, while MMU improves slightly with larger windows.}
\label{fig:future-window}
\end{figure}

To examine the sensitivity of VAM to the future window size $w_s$, we vary $w_s \in \{50, 100, 150, 200\}$ for each I2I channel and rebuild the value labels while keeping all other training and evaluation settings fixed. As shown in Figure~\ref{fig:future-window}, Swing and Marm achieve the best Recall@K at $w_s=100$ across cutoffs. Their performance at $w_s=50$ is consistently lower, and enlarging the window to 150 or 200 leads to only minor changes. On MMU, larger windows yield slightly higher recall, which indicates that the engagement attributed to retrieved items accumulates over a longer horizon for this channel. We use $w_s=100$ as the default window size in all subsequent experiments for a unified and stable setting.

\subsection{CATR for Cross-Channel Routing (RQ3)}

\begin{table}[t]
\centering
\caption{
Cross-channel uniqueness of retrieved candidates under joint multi-channel routing.
Uniq@K measures the fraction of channel-$c$ Top-$K$ results that are unique to this channel in the merged retrieval results.
U@K in the header is shorthand for Uniq@K at cutoff $K$.
}
\label{tab:uniq}
\renewcommand{\arraystretch}{1.05}
\begin{tabular}{l|lcccc}
\toprule
Channel & Method & U@100 & U@500 & U@1000 & U@2000 \\
\midrule
\multirow{2}{*}{Swing I2I}
  & \textit{w/o Div} & 0.8327 & 0.7918 & 0.7892 & 0.8180 \\
  & CAPTS   & 0.8895 & 0.8288 & 0.8115 & 0.8338 \\
\midrule
\multirow{2}{*}{Marm I2I}
  & \textit{w/o Div} & 0.8299 & 0.7883 & 0.7455 & 0.6553 \\
  & CAPTS   & 0.8895 & 0.8280 & 0.7827 & 0.6921 \\
\midrule
\multirow{2}{*}{MMU I2I}
  & \textit{w/o Div} & 0.9468 & 0.9406 & 0.9408 & 0.9481 \\
  & CAPTS   & 0.9595 & 0.9492 & 0.9460 & 0.9503 \\
\bottomrule
\end{tabular}
\end{table}

\subsubsection{Impact on Cross-Channel Uniqueness}

To verify whether the diversity objective in CATR improves the diversity of multi-channel retrieval results and reduces cross-channel redundancy, we evaluate three I2I channels under the joint routing setting. The relevant definitions are provided in Sec.~\ref{sec:capts_catr}. We use Uniq@K to measure each channel's unique contribution to the merged retrieval results. For a request and channel $c$, let $\mathcal{R}_c^K$ be the channel's Top-$K$ retrieved set and let $\mathcal{U}_c^K$ be the subset retrieved only by channel $c$. We compute
$\mathrm{Uniq}_c@K=\frac{|\mathcal{U}_c^K|}{|\mathcal{R}_c^K|+\varepsilon}$,
which is consistent with Eq.~\eqref{eq:catr_unique_set} and Eq.~\eqref{eq:catr_unique_ratio} when instantiating the retrieved set as Top-$K$ results. Table~\ref{tab:uniq} compares CAPTS with an ablated variant \textit{w/o Div} that removes the diversity objective while keeping other settings unchanged. The results show that CAPTS consistently increases Uniq@K across all channels and all reported $K$, indicating reduced cross-channel overlap and stronger complementarity in multi-channel retrieval.

\subsubsection{Routing Module Ablations}

\begin{table}[t]
\centering
\caption{
Routing module ablations under joint multi-channel routing.
\textit{w/o Div} removes the diversity objective and \textit{w/o Cal} removes the channel calibrator. Overall Recall@K is computed on the de-duplicated union of candidates from all channels.
}
\label{tab:ablation}
\renewcommand{\arraystretch}{1.05}
\begin{tabular}{l|lcccc}
\toprule
Channel & Method & R@100 & R@500 & R@1000 & R@2000 \\
\midrule
\multirow{3}{*}{Swing I2I}
  & \textit{w/o Div}  & 0.0162 & 0.0640 & 0.1081 & 0.1591 \\
  & \textit{w/o Cal} & 0.0151 & 0.0624 & 0.1065 & 0.1602 \\
  & CAPTS    & 0.0152 & 0.0627 & 0.1068 & 0.1610 \\
\midrule
\multirow{3}{*}{Marm I2I}
  & \textit{w/o Div}  & 0.0178 & 0.0678 & 0.1000 & 0.1018 \\
  & \textit{w/o Cal} & 0.0162 & 0.0657 & 0.1003 & 0.1026 \\
  & CAPTS    & 0.0163 & 0.0663 & 0.1016 & 0.1040 \\
\midrule
\multirow{3}{*}{MMU I2I}
  & \textit{w/o Div}  & 0.0042 & 0.0139 & 0.0215 & 0.0311 \\
  &\textit{w/o Cal} & 0.0041 & 0.0138 & 0.0215 & 0.0312 \\
  & CAPTS    & 0.0041 & 0.0138 & 0.0214 & 0.0311 \\
\midrule
\multirow{3}{*}{Overall}
  &\textit{w/o Div}  & 0.0262 & 0.0956 & 0.1495 & 0.1944 \\
  & \textit{w/o Cal} & 0.0253 & 0.0955 & 0.1507 & 0.1977 \\
  & CAPTS    & 0.0255 & 0.0963 & 0.1519 & 0.1994 \\
\bottomrule
\end{tabular}
\end{table}

Table~\ref{tab:ablation} compares CAPTS with two ablated variants, \textit{w/o Div} and \textit{w/o Cal}, under the joint multi-channel routing setting. We report Recall@K for each channel using its own retrieved candidates and also report Overall Recall@K on the de-duplicated union of candidates from all channels. Removing the channel calibrator in \textit{w/o Cal} leads to an overall drop across cutoffs. Concretely, Overall R@2000 decreases from 0.1994 to 0.1977, indicating that the channel calibrator provides a stable gain. In contrast, removing the diversity objective in \textit{w/o Div} can be slightly better than CAPTS for some channels at small cutoffs. Specifically, Swing increases from 0.0152 to 0.0162 at R@100 and Marm increases from 0.0163 to 0.0178 at R@100. However, it performs worse at larger cutoffs on the merged results, where Overall R@2000 drops from 0.1994 to 0.1944. This pattern indicates that optimizing per-channel Recall@K alone can improve performance at small cutoffs, but it increases cross-channel overlap and weakens the effective coverage of the merged candidate set. CATR learns trigger-to-channel routing and introduces the diversity objective to explicitly encourage cross-channel complementarity. Together with the channel calibrator for stabilizing value estimation, CATR improves the overall gains of multi-channel merged recall.

\subsection{Online Experiments (RQ4)}

\subsubsection{Online A/B Test Results}

\begin{table}[t]
\centering
\caption{
Online A/B test lifts.
Relative improvement over the strong baseline.
($p < 0.05$ for all metrics.)
}
\label{tab:online-ab}
\renewcommand{\arraystretch}{1.05}
\begin{tabular}{l|c}
\toprule
Metric & Relative improvement \\
\midrule
Daily Active Devices (DAD)         & +0.115\% \\
Total Time Spent (App)             & +0.713\% \\
Avg.\ Time Spent per Device        & +0.586\% \\
Total Watch Time (Video)           & +0.506\% \\
Avg.\ Watch Time per Device        & +0.395\% \\
\bottomrule
\end{tabular}
\end{table}

We deploy CAPTS in the production environment of Kwai and run large-scale online A/B tests.
Control and treatment use the same four production I2I channels---Swing~\cite{U2I2I_taobao}, Marm~\cite{Marm}, MMU~\cite{MMU}, and SimLR~\cite{SIM}---with identical retrieval, quotas, and downstream ranking. The only change is trigger selection and routing: control uses the incumbent policy and treatment uses CAPTS.
As shown in Table~\ref{tab:online-ab}, CAPTS improves overall engagement and also increases daily active devices (DAD) by +0.115\%.
Notably, CAPTS delivers a +0.586\% lift in average app time spent per device and a +0.713\% lift in total app time spent.
While the percentage appears small, it is a substantial gain for a mature retrieval stage, especially because the production change is confined to trigger selection and routing. This localized deployment pattern has also supported CAPTS deployment in Kwai's e-commerce recommendation scenario.

\subsubsection{Channel-wise Exposure and Quality Analysis}

\begin{figure}[t]
  \centering
  \includegraphics[width=\linewidth]{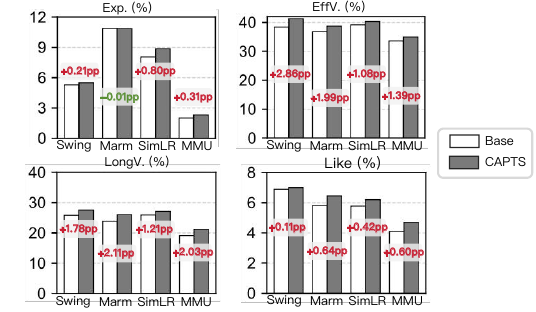}
  \caption{
    Channel-wise exposure share and engagement rates. We compare exposure share (Exp.), effective-view rate (EffV., watch time $\ge$ 7\,s), long-view rate (LongV.), and like rate (Like).
    }
  \Description{Grouped bar charts compare the rule-based baseline with CAPTS for four I2I channels. For each channel, CAPTS increases exposure share, effective-view rate, long-view rate, and like rate relative to the baseline.}
  \label{fig:online-channel}
\end{figure}

Figure~\ref{fig:online-channel} reports a post-hoc channel-wise breakdown of exposure and engagement for the four I2I retrieval paths. Overall, CAPTS increases the exposure contributed by I2I retrieval, where the summed exposure share of the four I2I channels rises from 26.24\% to 27.55\%, corresponding to a +1.31 percentage-point lift (pp). Meanwhile, channel-level engagement improves consistently across all four channels. Averaged over channels, CAPTS increases the effective-view rate, long-view rate, and like rate by approximately +1.83\,pp, +1.78\,pp, and +0.44\,pp, respectively. These results indicate that CAPTS not only makes I2I candidates more competitive in downstream ranking, but also improves conversion efficiency within each retrieval channel, leading to more reliable contributions from multi-channel I2I retrieval in production.

\section{Related Work}\label{sec:related_works}

\subsection{Retrieval in Recommender Systems}
Retrieval underpins candidate generation over massive corpora, and industrial systems adopt two paradigms: direct user-to-item retrieval (U2I) and two-hop user-to-item-to-item retrieval (U2I2I). \textbf{Direct Embedding Retrieval (U2I)} maps users and items into a shared latent space for nearest neighbor search~\cite{youtube, DSSM}. To represent polysemous interests, methods such as MIND~\cite{MIND} and ComiRec~\cite{ComiRec} use routing or attention, and recent work improves representations with contrastive learning (e.g., CL4SRec~\cite{CL4SRec}, ICL~\cite{ICL}) or joint optimization (e.g., Uni-Retriever~\cite{Uni-Retriever}), but U2I can be less controllable than multi-step retrieval. \textbf{Two-hop Retrieval (U2I2I)} instead selects a small set of historical items as triggers and expands them via item-to-item retrieval, and is widely deployed in systems such as Pinterest~\cite{PinSage, Pixie}, Alibaba~\cite{U2I2I_taobao}, and Twitter~\cite{TwHIN}. Beyond rule-based trigger choices, PDN~\cite{PDN} models user$\rightarrow$trigger$\rightarrow$item paths to estimate trigger importance and I2I relevance, selecting top-$m$ triggers and retrieving top-$k$ items per trigger, yet trigger selection is often optimized for short-horizon signals and a single retrieval path, leaving preference-aligned value attribution and trigger routing across heterogeneous multi-channel I2I retrieval underexplored.

\subsection{Interest Selection and Trigger Learning}
Selecting behaviors from user sequences matters for preference modeling, and existing methods can be grouped by whether the target item is available when weighting histories. \textbf{Target-Aware Sequence Modeling} assumes a ranking candidate and uses attention to compute target-conditioned weights over behaviors (e.g., DIN~\cite{DIN}, DIEN~\cite{DIEN}); for long sequences, SIM~\cite{SIM} and SDIM~\cite{SDIM} use multi-stage retrieval or hash lookup, and HSD~\cite{HSD} applies denoising to filter irrelevant interactions. However, these methods rely on a target and thus do not apply directly to trigger selection, where downstream candidates are unknown and the system must select a small trigger set under strict budgets. \textbf{Target-Agnostic Trigger Learning} selects retrieval seeds without an explicit target and remains less explored in practice, where systems still rely on rule-based triggers such as recency or frequency. PDN~\cite{PDN} learns trigger importance for two-hop retrieval, while Trinity~\cite{Trinity} synthesizes multiple interests with rule-based and statistical designs to mitigate interest amnesia. Despite these advances, trigger selection is often optimized with short-horizon objectives and a single retrieval path, rather than attributing value by future engagement induced by the recalled set, and it typically does not model trigger-to-channel allocation under heterogeneous multi-channel retrieval; CAPTS addresses these gaps with look-ahead value attribution and channel-adaptive routing.

\section{Conclusion}\label{sec:conclusion}
We studied trigger selection for multi-channel item-to-item (I2I) retrieval and proposed CAPTS, a unified framework that aligns trigger scoring with look-ahead retrieval utility and performs channel-aware routing under per-channel budgets. CAPTS combines a Value Attribution Module (VAM) that derives per-trigger, per-channel supervision from time-consistent I2I replay and future-window engagement with a Channel-Adaptive Trigger Routing (CATR) model that learns calibrated per-channel value predictions and encourages cross-channel complementarity via a diversity objective. By localizing optimization to trigger selection, CAPTS can improve multi-channel U2I2I retrieval while keeping existing I2I retrievers and downstream rankers intact. Its successful extension to Kwai's e-commerce recommendation scenario further suggests practical cross-scenario portability. The current formulation uses fixed per-channel budgets and replay-derived supervision; future work could jointly adapt channel budgets and model engagement beyond the observation window. Extensive offline evaluations and large-scale online A/B tests on a short-video platform show that CAPTS consistently improves multi-channel recall and yields measurable production gains, including +0.713\% total app time spent and +0.586\% app duration per device.

\bibliographystyle{ACM-Reference-Format}
\bibliography{references}

\appendix

\section{Notation}\label{app:notation}

Table~\ref{tab:notation} summarizes the notation used in the formulation of
VAM and CATR.

\begin{table}[H]
\centering
\caption{Summary of the main notation.}
\label{tab:notation}
\small
\setlength{\tabcolsep}{3pt}
\renewcommand{\arraystretch}{1.04}
\begin{tabular}{@{}p{0.27\columnwidth}p{0.67\columnwidth}@{}}
\toprule
Symbol & Description \\
\midrule
$u$, $i$ & User and item (video), respectively. \\
$\mathcal{H}_u$ & Interaction history of user $u$. \\
$(i_k,\tau_k,d_k,r_k)$ & Item, timestamp, watch time, and explicit feedback in the $k$-th interaction. \\
$\tau_0$ & Timestamp of the current request. \\
$\mathcal{C}$, $c$ & Set of I2I channels and a channel index. \\
$B_c$ & Per-request trigger budget for channel $c$. \\
$t$ & Candidate trigger item. \\
$\mathcal{T}_u(\tau_0)$ & Eligible triggers observed before $\tau_0$. \\
$\mathcal{W}_u^{\mathrm{fut}}(\tau_0)$ & Items consumed in the next $w_s$ effective views. \\
$w_s$ & Future-window size used by VAM. \\
$\mathcal{R}_c(t,\tau_0)$ & Items retrieved by channel $c$ from trigger $t$ at $\tau_0$. \\
$g(j)$ & Engagement observed on retrieved item $j$. \\
$V_c(t)$ & Forward value of routing $t$ to channel $c$. \\
$\ell_c(t,\tau_0)$ & Clipped channel-specific engagement intensity. \\
$s_c$, $M_c$ & Scale factor and clipping cap for the intensity label. \\
$y_c(t,\tau_0)$ & Binary channel-specific value label. \\
$\mathbf{h}_u$, $\mathbf{h}_t$ & User and trigger representations in CATR. \\
$\tilde{\mathbf{h}}_t$ & Trigger-aware behavior summary from target attention. \\
$\hat{v}_c(t)$ & Calibrated value prediction for trigger $t$ on channel $c$. \\
$\rho_c(t,\tau_0)$ & Fraction of results unique to channel $c$ for trigger $t$. \\
$\eta$ & Diversity weight in the routing score. \\
$\mathrm{Uniq}_c@K$ & Fraction of channel-$c$ Top-$K$ results unique to that channel. \\
\bottomrule
\end{tabular}
\end{table}

\section{Implementation Details}\label{app:impl}

\paragraph{Feature processing and supervision.}
For each logged request, we replay every candidate trigger on each I2I channel
and construct the VAM targets according to
Eqs.~\eqref{eq:vam_reward}--\eqref{eq:vam_binary_label}. CATR encodes the
most recent $L=50$ effective views as the behavior sequence $\mathbf{X}$. We
use $s_c=100$ and $M_c=6$ as the default channel-wise scale and clipping cap,
respectively. Time-related continuous features, including watch time and time
gaps, are discretized into buckets to reduce sensitivity to heavy-tailed
values. The future-window size is $w_s=100$ unless otherwise stated.

\paragraph{Retrieval channels.}
The offline experiments replay three channels that represent the principal
retrieval paradigms in the production system. The online A/B test additionally
enables SimLR I2I, resulting in the four-channel configuration described below.

\begin{itemize}[leftmargin=*,nosep]
  \item \textbf{Swing I2I}~\cite{U2I2I_taobao} is a collaborative-filtering
  channel based on co-occurrence signals from user interactions.
  \item \textbf{Marm I2I}~\cite{Marm} is a deep embedding-matching channel
  designed to capture long-term user interests.
  \item \textbf{MMU I2I}~\cite{MMU} is a multi-modal content channel that
  retrieves videos using visual and semantic similarity.
  \item \textbf{SimLR I2I}~\cite{SIM} is a pairwise logistic-regression
  channel that learns item--item similarity from combined user--item features.
\end{itemize}

\section{Sensitivity to the Diversity Weight}\label{app:eta-sensitivity}

The diversity weight $\eta$ in Eq.~\eqref{eq:catr_route_score} controls the
trade-off between channel-specific value and cross-channel complementarity.
Table~\ref{tab:eta-sensitivity} shows that increasing $\eta$ from 0 to 0.4
substantially improves R@2000, while R@100 decreases slightly. Performance at
R@2000 plateaus for $\eta\in\{0.4,0.5\}$ and declines when $\eta$ is
increased to 0.6. We therefore use $\eta=0.4$, which attains the best R@2000
while placing the smaller weight on the diversity term among the tied settings.

\begin{table}[H]
\centering
\caption{Sensitivity of overall Recall@K to the diversity weight $\eta$.}
\label{tab:eta-sensitivity}
\renewcommand{\arraystretch}{1.05}
\begin{tabular}{c|cc}
\toprule
$\eta$ & R@100 & R@2000 \\
\midrule
0.0 & 0.0262 & 0.1944 \\
0.2 & 0.0257 & 0.1985 \\
0.4 & 0.0255 & \textbf{0.1994} \\
0.5 & 0.0253 & \textbf{0.1994} \\
0.6 & 0.0251 & 0.1990 \\
\bottomrule
\end{tabular}
\end{table}

\section{Timestamp-Aligned I2I Replay}\label{app:timestamp}

VAM constructs supervision by attributing subsequent consumption to triggers
through I2I replay anchored at request time $\tau_0$. Because the replay runs
after $\tau_0$ while I2I indexes and retrieval models are continuously
refreshed, querying the latest artifacts could expose updates that were not
available at request time and introduce temporal leakage. We therefore use an
artifact snapshot aligned with $\tau_0$, so that $\mathcal{R}_c(t,\tau_0)$
approximates the retrieval state seen by the online request. Each channel
returns its Top-50 items per trigger, matching the online configuration.

Retrieval artifacts can be refreshed before the corresponding interaction
stream becomes available for labeling. To absorb this asynchronous lag, we use
a conservative cutoff $\tau_0^-=\tau_0-\delta$ and query only artifacts no
later than that cutoff. In our deployment, $\delta$ is approximately 20
minutes, reflecting the observed lead of retrieval refresh over interaction
stream availability. This rollback preserves request-time consistency without
changing the online retrievers used for evaluation or serving.

\section{Cross-Channel Analysis of Selected Triggers}\label{app:online-cross}

Table~\ref{tab:crosschain} provides a post-hoc comparison between CAPTS and
the rule-based control in the online experiment. Only 5.9\%--12.1\% of the
CAPTS-selected triggers also appear in the corresponding control sets,
confirming that CAPTS induces a materially different routing policy. Across all
four channels, the CAPTS-attributed impressions have higher average play time
(+5.4\% to +14.9\%) and average video duration (+2.2\% to +13.5\%). The mean
absolute gains are 2.29 seconds and 6.18 seconds, respectively. Level-1 category
coverage also increases on Swing, SimLR, and MMU, by one category on average
across the four channels. These diagnostic results are consistent with the
mechanism targeted by CATR: routing distinct triggers toward channel-specific,
complementary retrieval results. As a post-hoc analysis, they provide
supporting evidence rather than a separate causal estimate of each mechanism.

\begin{table}[H]
\centering
\caption{Cross-channel trigger overlap and channel-attributed item statistics
in the online A/B test. Overlap is measured against the rule-based control;
Improv.\ is the absolute CAPTS--control difference.}
\label{tab:crosschain}
\small
\setlength{\tabcolsep}{2.2pt}
\renewcommand{\arraystretch}{1.03}
\begin{tabular}{@{}l|lrrrr@{}}
\toprule
\multirow{2}{*}{Channel} & \multirow{2}{*}{Group} & Overlap & Avg.\ Play & Avg.\ Dur. & L1 Cat. \\
                         &                         & Rate    & (s)        & (s)       & Count \\
\midrule
\multirow{3}{*}{Swing I2I}
  & Control & --     & 26.22 & 91.96 & 25 \\
  & CAPTS   & 10.3\% & 27.63 & 94.02 & 26 \\
  & Improv. & --     & +1.41 & +2.06 & +1 \\
\midrule
\multirow{3}{*}{Marm I2I}
  & Control & --     & 23.87 & 79.83 & 28 \\
  & CAPTS   & 12.1\% & 27.43 & 90.62 & 28 \\
  & Improv. & --     & +3.56 & +10.79 & 0 \\
\midrule
\multirow{3}{*}{SimLR I2I}
  & Control & --     & 25.50 & 81.93 & 21 \\
  & CAPTS   & 8.7\%  & 27.66 & 87.46 & 23 \\
  & Improv. & --     & +2.16 & +5.54 & +2 \\
\midrule
\multirow{3}{*}{MMU I2I}
  & Control & --     & 27.25 & 91.33 & 27 \\
  & CAPTS   & 5.9\%  & 29.29 & 97.64 & 28 \\
  & Improv. & --     & +2.04 & +6.31 & +1 \\
\bottomrule
\end{tabular}
\end{table}

\end{document}